\documentclass[11pt]{article}
\usepackage{graphicx}

\newcommand{\BABARPubYear}    {04}

\newcommand{\BABARConfNumber} {002}
\newcommand{\SLACPubNumber} {10597}
\newcommand{\LANLNumber} {0000}

\newcommand{\xf}{\mbox{${\cal F}$}}
\newcommand{\costhr}{\ensuremath{\cos\theta_{\rm T}}}

\def\Bu      {\ensuremath{B^+}}
\def\Bub     {\ensuremath{B^-}}
\def\Bbar    {\overline{B}{}}

\def\Bzb     {\ensuremath{\Bbar^0}}
\def\Bz      {\ensuremath{B^0}}
\def\BpBm    {\ensuremath{\Bu  \Bub}}
\def\BzBzb   {\ensuremath{\Bz  \Bzb}}
\newcommand{\DE}{\ensuremath{\Delta E}}
\newcommand{\pvec}{{\bf p}}
\newcommand{\half}{\mbox{${1\over2}$}}
 \def\mes{\mbox{$m_{\rm ES}$}}
\newcommand{\calB}{\mbox{${\cal B}$}}
\newcommand{\auno}{\mbox{$a_1(1260)$}}
\newcommand{\aunob}{\mbox{$a_1$}}
\newcommand{\aunop}{\mbox{$a^+_1(1260)$}}

\newcommand{\appim}{\mbox{$a^+_1(1260) \, \pi^-$}}
\newcommand{\btoappim}{\mbox{$B^0 \rightarrow a^+_1(1260)\, \pi^-  $}}

\newcommand{\UfourS}{\mbox{$\Upsilon(4S)$}}
\newcommand{\Brapi}{\mbox{$\calB(\btoappim)$}}
\newcommand{\rapi}{\mbox{$42.6 \pm 4.2  \pm 4.1$}}
\newcommand{\Rapi}{\mbox{$(\rapi)\times 10^{-6}$}}
\newcommand{\epem}{\mbox{$e^+e^-$}}
\def\babar{{\em B}{\footnotesize\em A}{\em B}{\footnotesize\em AR}}
\def\BB{\mbox{$B\overline B\ $}}
\def\pep2{PEP-II}

\newcommand{\Dcontrol}{\mbox{$B \rightarrow \pi^- D^0 , D^0 \rightarrow K^{-} \pi^+ \pi^0$}}
\newcommand\etal{{\it et al.}}
\newcommand{\dedx}{\ensuremath{\mathrm{d}\hspace{-0.1em}E/\mathrm{d}x}}
\newcommand{\gevcc}{\mbox{$\textrm{GeV}/c^2$}} 
\newcommand{\mevcc}{\mbox{$\textrm{MeV}/c^2$}} 
\newcommand{\gevc}{\mbox{$\textrm{GeV}/c$}} 

\newcommand{\gev}{\mbox{$\textrm{GeV}$}} 
\newcommand{\mev}{\mbox{$\textrm{MeV}$}} 
\newcommand{\mrad}{\mbox{$\textrm{mrad}$}} 

\newcommand{\jprlBase}  [1]     {Phys.\ Rev.\ Lett.}
\newcommand{\jprl}      [1]    {\jprlBase\ B~{\bf #1}}

\newcommand{\jprBase}        {Phys.\ Rev.\ }
\newcommand{\jprd}      [1]  {\jprBase\ D~{\bf #1}}
\newcommand{\plBase}   [1]         {Phys.\ Lett.}
\newcommand{\plb}      [1]    {\plBase\ B~{\bf #1}}

\newcommand{\nimBaseA}       {Nucl.\ Instr.\ Meth.\ }
\newcommand{\nima}      [1]  {\nimBaseA~A~{\bf #1}}

\newcommand{\npBase}         {Nucl.\ Phys.\ }
\newcommand{\npb}       [1]  {\npBase\ B~{\bf #1}}

\long\def\inst#1{\par\nobreak\kern 4pt\nobreak
    {\it #1}\par\vskip 10pt plus 3pt minus 3pt}

\begin{document}
{\pagestyle{empty}

\begin{flushright}
\babar-CONF-\BABARPubYear/\BABARConfNumber \\
SLAC-PUB-\SLACPubNumber \\
hep-ex/\LANLNumber \\
August 2004 \\
\end{flushright}

\par\vskip 5cm

\begin{center}
\Large \bf Observation of $B^0$ Meson Decay to \appim
\end{center}
\bigskip

\begin{center}
\large The \babar\ Collaboration\\
\mbox{ }\\
\today
\end{center}
\bigskip \bigskip

\begin{center}
\large \bf Abstract
\end{center}

We present a preliminary  measurement of the branching fraction of the $B$ meson decay \\
\btoappim\   with \aunop\  $\rightarrow \pi^+ \pi^+ \pi^-$. The data were recorded with the \babar\ detector 
at the SLAC $B$ factory \pep2  and correspond to 
$124\times 10^6$ \BB\ pairs produced in \epem\ annihilation through the
\UfourS\ resonance.  We find the branching fraction \Brapi = \Rapi.
The fitted values of the \auno\ parameters are $m_{\aunob }=1.19 \pm 0.02$ \gevcc\ and
$\Gamma_{\aunob } = 312 \pm 55$ \mevcc.

\vfill
\begin{center}

Submitted to the 32$^{\rm nd}$ International Conference on High-Energy Physics, ICHEP 04,\\
16 August---22 August 2004, Beijing, China

\end{center}

\vspace{1.0cm}
\begin{center}
{\em Stanford Linear Accelerator Center, Stanford University, 
Stanford, CA 94309} \\ \vspace{0.1cm}\hrule\vspace{0.1cm}
Work supported in part by Department of Energy contract DE-AC03-76SF00515.
\end{center}

\newpage
} 

\begin{center}
\small

The \babar\ Collaboration,
\bigskip

%
B.~Aubert,
R.~Barate,
D.~Boutigny,
F.~Couderc,
J.-M.~Gaillard,
A.~Hicheur,
Y.~Karyotakis,
J.~P.~Lees,
V.~Tisserand,
A.~Zghiche
\inst{Laboratoire de Physique des Particules, F-74941 Annecy-le-Vieux, France }
% ...

\end{center}\newpage

\section{INTRODUCTION}

We report on the preliminary  measurement of  the branching fraction
$B^0 \rightarrow  a_1^+(1260) \pi^-$ with
$a_1^+(1260) \rightarrow \pi^+ \pi^+ \pi^-$ \cite{Coniugati} .
The  $a_1(1260) \rightarrow 3 \pi$
decay  proceeds mainly through the intermediate states
$(\pi \pi)_{\rho} \pi$ and $(\pi \pi)_{\sigma} \pi$.

The study of this decay mode is complicated by open questions on the
parameters of the $a_1(1260)$ meson. There are large discrepancies between
 these
parameters when comparing results from analyses involving hadronic interactions
~\cite{palombo}
and $\tau$ decays~\cite{tau}. Therefore, it is important
to verify the theoretical prediction of the branching fraction for
this decay mode and have  new measurements of the $a_1(1260)$ parameters.
It is also important to note that the $B^0 \rightarrow  a_1^+(1260) \pi^-$ channel
can be used to measure the Cabibbo-Kobayashi-Maskawa
angle $\alpha$ of the Unitarity triangle~\cite{aleksan}.

There has been no experimental observation of this decay mode. An upper
limit of $49 \times 10^{-5}$ at the 90\% C.L. has been set by CLEO~\cite{CLEO}
for the branching fraction of $B^0 \rightarrow a_1^+(1260) \pi^-$,
while the DELPHI~\cite{Delphi} collaboration
has set the 90\% C.L. upper limit of $28 \times 10^{-5}$ for the
branching fraction of $B^0 \rightarrow 4\pi$.

Below we present the details of the analysis for the measurement of the
branching fraction for $B^0 \rightarrow  a_1^+(1260) \pi^- \rightarrow 4 \pi$.
Presently, we do not distinguish between the final
states $(\pi \pi)_{\rho} \pi$ and $(\pi \pi)_{\sigma} \pi$. Such an analysis
would require a study of the angular distributions of the decay products. 
Background contributions from $B^0$ decays to $a_2(1320) \pi$ and $\pi(1300) \pi$
were assumed to be negligible.

\section{THE \babar\ DETECTOR AND DATASET}
\label{sec:babar}

The results presented in this paper are based on data collected
in 1999--2003 with the \babar\ detector~\cite{BABARNIM}
at the PEP-II asymmetric $e^+e^-$ collider~\cite{pep}
located at the Stanford Linear Accelerator Center.  An integrated
luminosity of 112~fb$^{-1}$, corresponding to
124 million \BB\ pairs, was recorded at the $\Upsilon (4S)$
resonance
(``on-resonance'', center-of-mass energy $\sqrt{s}=10.58\ \gev$).
An additional 12~fb$^{-1}$ were taken about 40~MeV below
this energy (``off-resonance'') for the study of continuum background in
which a light or charm quark pair is produced instead of an \UfourS.

The asymmetric beam configuration in the laboratory frame
provides a boost of $\beta\gamma = 0.56$ to the $\Upsilon(4S)$.
Charged particles are detected and their momenta measured by the
combination of a silicon vertex tracker (SVT), consisting of five layers
of double-sided silicon detectors, and a 40-layer central drift chamber,
both operating in the 1.5-T magnetic field of a solenoid.
The tracking system covers 92\% of the solid angle in the CM frame.

Charged-particle identification (PID) is provided by the average
energy loss (\dedx\ ) in the tracking devices  and
by an internally reflecting ring-imaging
Cherenkov detector (DIRC) covering the central region.
A $K/\pi$ separation of better than four standard deviations ($\sigma$)
is achieved for momenta below 3 \gevc , decreasing to 2.5 $\sigma$ at the highest
momenta in the $B$ decay final states.
Photons and electrons are detected by a CsI(Tl) electromagnetic calorimeter (EMC).
The EMC provides good energy and angular resolutions for detection of photons in
 the range from 30 \mev\ to 4 \gev . The energy and angular resolutions are 3\% and 4 \mrad\ , respectively , for a 1 \gev\ photon.

The flux return for the solenoid is composed of multiple layers of iron
and resistive plate chambers for the identification of muons and long-lived
neutral hadrons.

\section{ANALYSIS METHOD}
\label{sec:Analysis}

Monte Carlo (MC) simulations \cite{geant4}   of the signal decay modes and
of continuum and \BB\ backgrounds are used to establish the event selection
criteria. We select \aunop\  candidates with the
following requirement on the invariant mass in \gevcc\ : $0.6 < m_{\auno}<1.8$ .
 The intermediate dipion state is reconstructed
with an invariant  mass between  0.46 and 1.1 \gevcc .

We make several particle identification requirements to ensure the
identity of the signal pions. For the bachelor
charged track we require an associated DIRC Cherenkov angle between
$-2\,\sigma$ and $+5\,\sigma$ from the expected value for a pion.

A $B$ meson candidate is characterized kinematically by the energy-substituted 
mass $\mes = \sqrt{(\half s + \pvec_0\cdot \pvec_B)^2/E_0^2 - \pvec_B^2}$ and
energy difference $\DE = E_B^*-\half\sqrt{s}$, where the subscripts $0$ and
$B$ refer to the initial \UfourS\ and to the $B$ candidate  in the lab-frame, respectively,
and the asterisk denotes the \UfourS\ frame.
We require $|\DE|\le0.2$ GeV and $5.25\le\mes\le5.29\ \gevcc$.  The momentum of \aunop\ in the
center-of-mass frame is required to be between 2.3 and 2.7 \gevc. To reduce fake $B$ meson
candidates we require p($\chi^2$) $>$ 0.01 for the $B$ vertex fit. The angular
variable $\mathcal{H}_{a_1}$ (cosine of the angle between the direction of the
$\pi$ meson  with respect to the flight direction of the $B$  in the \auno\ meson rest frame)
is required to be between -0.85 and 0.85 to suppress combinatorics.

To reject continuum background, we make use
of the angle $\theta_T$ between the thrust axis of the $B$ candidate and
that of
the rest of the tracks and neutral clusters in the event, calculated in
the center-of-mass frame.  The distribution of $\cos{\theta_T}$ is
sharply peaked near $\pm1$ for combinations drawn from jet-like $q\bar q$
pairs and is nearly uniform for the isotropic $B$ meson decays; we require
$|\cos{\theta_T}|<0.65$.  
The remaining continuum background is modeled from sideband data. 
We use Monte Carlo simulations of \BzBzb\ and \BpBm\ decays
to look for \BB\ backgrounds, which can come from both charmless 
and charm decays. 
We find that the decay mode $B^0 \rightarrow D^- \pi^+$, with 
$D^- \rightarrow K^+ \pi^- \pi^-$ and $D^- \rightarrow K^0_S \pi^-$, is the only
significant background, and is included in the maximum likelihood fit. 

We use an unbinned, multivariate maximum-likelihood fit to extract
the signal yield for $B^0 \rightarrow  a_1^+(1260) \pi^-$.
The likelihood function incorporates four uncorrelated variables.
We describe the $B$ decay kinematics with two variables: \DE\ and \mes, as
mentioned above. We also include $m_{a_1}$ and a
Fisher discriminant \xf\ which describes the energy flow in the event.
The Fisher discriminant combines four
variables: the angles with respect to the beam axis, in the \UfourS\
frame, of the $B$ momentum and $B$ thrust axis, and
the zeroth and second angular moments $L_{0,2}$ of the energy flow
around the $B$ thrust axis.  The moments are defined by
\begin{equation}
  L_j = \sum_i p_i\times\left|\cos\theta_i\right|^j,
\end{equation}
where $\theta_i$ is the angle with respect to the $B$ thrust axis of
track or neutral cluster $i$, $p_i$ is its momentum, and the sum
excludes tracks and clusters used to build the $B$ candidate.

Since the correlations between the observables in the selected data
are small, we take the
probability density function (PDF) for each event to be a product
of the PDFs for the separate observables.
The product PDF for event $i$ and hypothesis $j$, where
$j$ can be signal, continuum background or \BB\ background, is given by

\begin{equation}
{\cal P}^i_{j} =  {\cal P}_j (\mes) \cdot {\cal  P}_j (\DE) \cdot
 { \cal P}_j(\xf) \cdot {\cal P}_j (m_{a_1}).
\end{equation}

There is the possibility that a track from a signal
event is exchanged with a track from the rest of the event. We call these events
``self-cross-feed'' (SCF) events. The fraction
of SCF events with respect to the total number of signal events, $f_{SCF}$,
is found to be 0.31 from Monte Carlo studies.

\noindent The likelihood function for the event $i$ is defined as :
\begin{equation}
{\cal L}^{i} = n_{sig}(1-f_{SCF}){\cal P}_{sig}^{i}+n_{sig}f_{SCF}
                 {\cal P}_{SCF}^{i}+n_{q\bar{q}}{\cal P}_{q\bar{q}}^{i} +
                 n_{B\bar{B}1}{\cal P}_{B\bar{B}1}^{i}+ n_{B\bar{B}2}{\cal P}_{B
\bar{B}2}^{i}\,,
 \end{equation}

\noindent where $ n_{sig}$ is the number of signal events,  $n_{q\bar{q}}$ the
number of continuum
 background events, $n_{B\bar{B}1}$ the number of \BB\ background events
$D^-\pi^+$ with $K^+ \pi^- \pi^-$
and  $n_{B\bar{B}2}$ the number of \BB\ background events $D^-\pi^+$ with $K^0_
S \pi^-$. The extended likelihood function for all events is :
\begin{equation}
{\cal L} = \frac{\exp{(-\sum_j n_{j})}}{N!}\prod_i^{N}\sum_j n_{j} {\cal P}^i_{j
}\,,
\end{equation}

\noindent where $n_{j}$ is the yield of events of hypothesis $j$ found by the
fitter, and $N$ is the number of events in the sample.  The first factor takes
into account the Poisson fluctuations in the total number of events.

We determine the PDFs for signal and \BB\ backgrounds from
MC distributions in each observable.  For the continuum background we establish
the functional forms and initial parameter values of the PDFs with data
from sidebands in \mes\ or \DE.  We allow the signal  $m_{a_1}$ PDF parameters and
several background PDF parameters to float in the final fit.

The distribution  of $m_{a_1}$ in signal events  is parameterized as a relativistic Breit-Wigner.
The \mes\ and \DE\ distributions for signal are parameterized as
double Gaussian functions. Slowly varying distributions are
parameterized by linear functions.
The combinatoric background in \mes\ is described by a phase-space-motivated
empirical function \cite{argus}.  We model the
\xf\ distribution using a Gaussian function with different widths above
and below the mean.

Possible differences between Monte Carlo simulation and on-resonance data are investigated using
the control sample \Dcontrol, which has a similar topology to the signal mode.

\section{SYSTEMATIC STUDIES}
\label{sec:Systematics}

Most of the systematic errors on the yields that arise from uncertainties in the
values of the PDF parameters have already been incorporated into the overall
statistical error, since they are floated in the fit.  
We determine the sensitivity to the other parameters of the signal PDF
components by varying these within their uncertainties.  The results are
shown in the first row of Table \ref{tab:systtab}.  This is the only
systematic error on the fit yield; the other systematics apply to either
the efficiency or the number of \BB pairs in the data sample.

The uncertainty in our knowledge of the efficiency is found
to be 0.8$N_t$\%, where $N_t$ is the number of signal tracks.
We estimate the uncertainty in the number of
\BB pairs to be 1.1\%. The fitting algorithm introduces 
a systematic bias of 3.1\%, which was found from fits to
simulated samples with varying background populations.  Published world
averages \cite{PDG2002}\ provide the $B$ daughter branching fraction
uncertainties. The systematic error from $a_1(1260) K$ cross-feed background
is estimated to be 5\%, while the systematic error due to SCF is found to be 3\%.
A systematic error of 1\% is assigned to potential contributions from $B \rightarrow 4\pi$ and
$B \rightarrow \rho \pi\pi$. Finally, we account for systematic effects in \costhr\ (1\%) and in the PID
requirement (0.5\%) on the prompt charged track. The values for each of
these contributions are given in Table \ref{tab:systtab}.

\begin{table}[htbp]
\caption{Estimates of the systematic errors (in percent).}
\label{tab:systtab}
\begin{center}
\begin{tabular}{l|c}
\hline\hline
Quantity & $a_1^+ \pi^- $\\
\hline
Fit yield              &$6.3$   \\
Fit eff/bias           &$1.7$  \\
Track multiplicity     & $1.0$ \\
Tracking eff/qual     & $3.2$ \\
Number \BB\            & $1.1$\\
SCF                    & $3$ \\
$a_1 K$ cross-feed     & $5$ \\
$B  \rightarrow 4 \pi$,$\rho \pi \pi$ & $1.0 $  \\
MC statistics          & $1.0$ \\
\costhr                 & $1.0$ \\
\hline
Total                   & $9.6$  \\
\hline\hline
\end{tabular}
\end{center}
\end{table}

\section{RESULTS}
\label{sec:Physics}

By generating (from the PDFs) and fitting simulated samples of signal
and background, we verify that our fitting procedure is working
properly.  We find that the minimum $\ln{\cal L}$ value for the
on-resonance data lies well within the $\ln{\cal L}$ distribution
from these simulated samples.

The efficiency is obtained from the fraction of signal MC events passing
the selection criteria, adjusted for any bias in the likelihood fit.  This bias is
determined from fits to simulated samples, each equal in size to the
data and containing a known number of signal MC events combined with events
generated from the background PDFs.  We find a fit bias of 0.97.

\begin{table}[htb]
\caption{
Final fit results.
}
\label{tab:results}
\begin{center}
\vspace*{-0.2cm}
\hspace*{-1.0cm}
\begin{tabular}{l|c}
\hline\hline
Fit quantity    &\appim\  \\
\hline
Fit sample size         &  \\
~~On-resonance          & 32500\\
~~Off-resonance         & 2680 \\
Signal yield            &     \\
~~On-res data           &$472.3^{+46.8}_{-45.9}$ \\
~~Off-res data          &$6.2^{+10.8}_{-8.4}$  \\
Selection $\epsilon$ (\%)& $19.4$        \\
Track corr.              & $0.953$  \\
Fit-bias                 & $0.967$ \\
$\prod\calB_i$ (\%)      & $50$  \\
\hline
Stat. sign. ($\sigma$)   & $13.8$    \\
${\cal B}(\times 10^{-6})$      &$42.6 \pm 4.2 \pm 4.1$  \\
\hline\hline\\
\end{tabular}
\end{center}
\end{table}

In Table~\ref{tab:results} we show the results of the fits for
on- and off-resonance data.
We also show the fitted signal yield, the efficiency ($\epsilon$), the 
daughter branching fraction product ($\prod\calB_i$), the statistical significance, and the
central value of the branching fraction.
The statistical error on the number of events
is taken to be the change in the central value when the quantity
$-2\ln{\cal L}$ changes by one unit. The statistical significance is
taken as the square root of the difference between the value of
$-2\ln{\cal L}$ for zero signal and the value at its minimum.

\begin{figure}[h]

 \begin{minipage}{\linewidth}
  \begin{center}
\vspace{1.0cm}
   \includegraphics[bb=85 155 535 605 ,angle=270,scale=0.30]{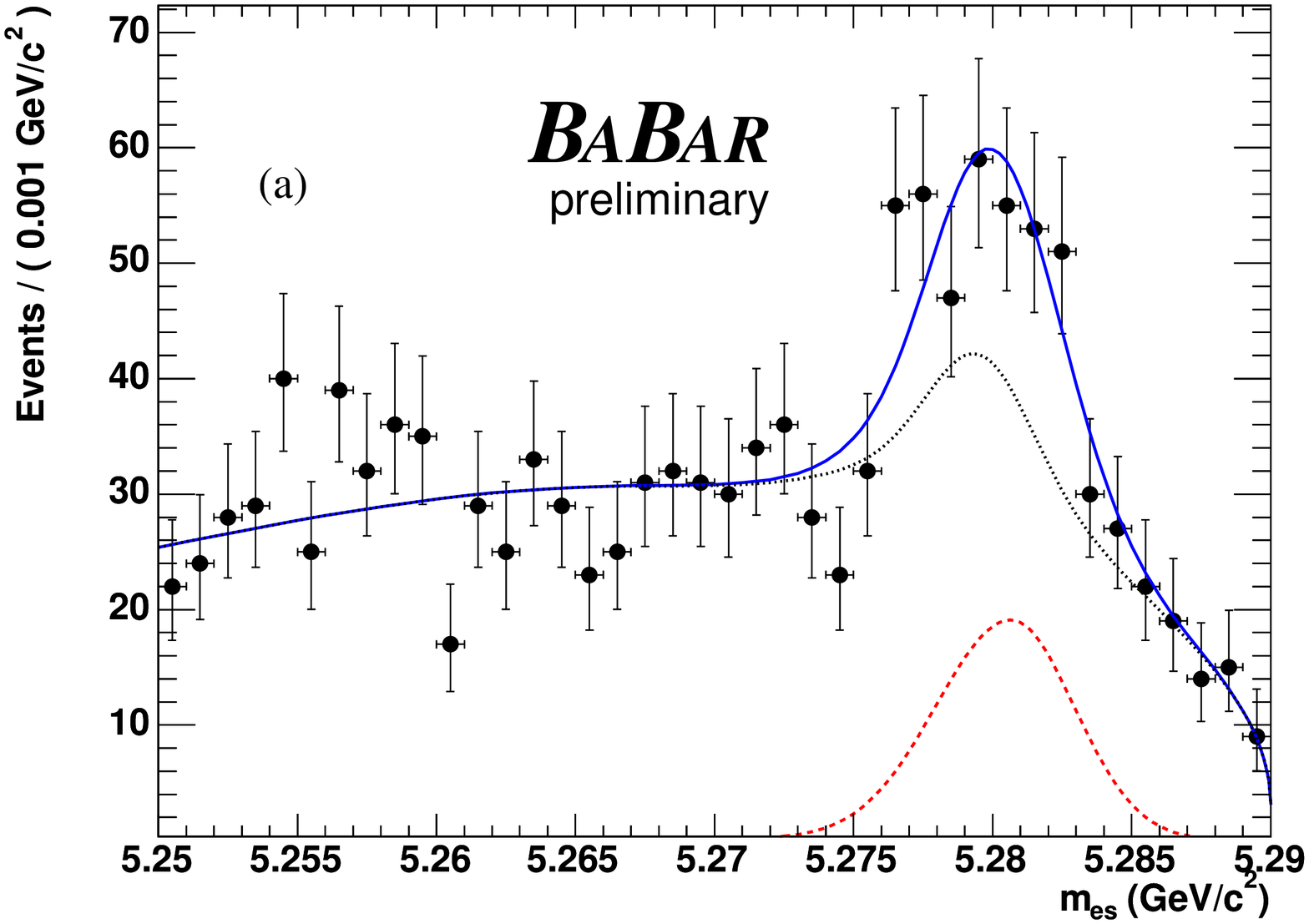}
   \hspace{3cm}
   \includegraphics[bb=85 155 535 605 ,angle=270,scale=0.30]{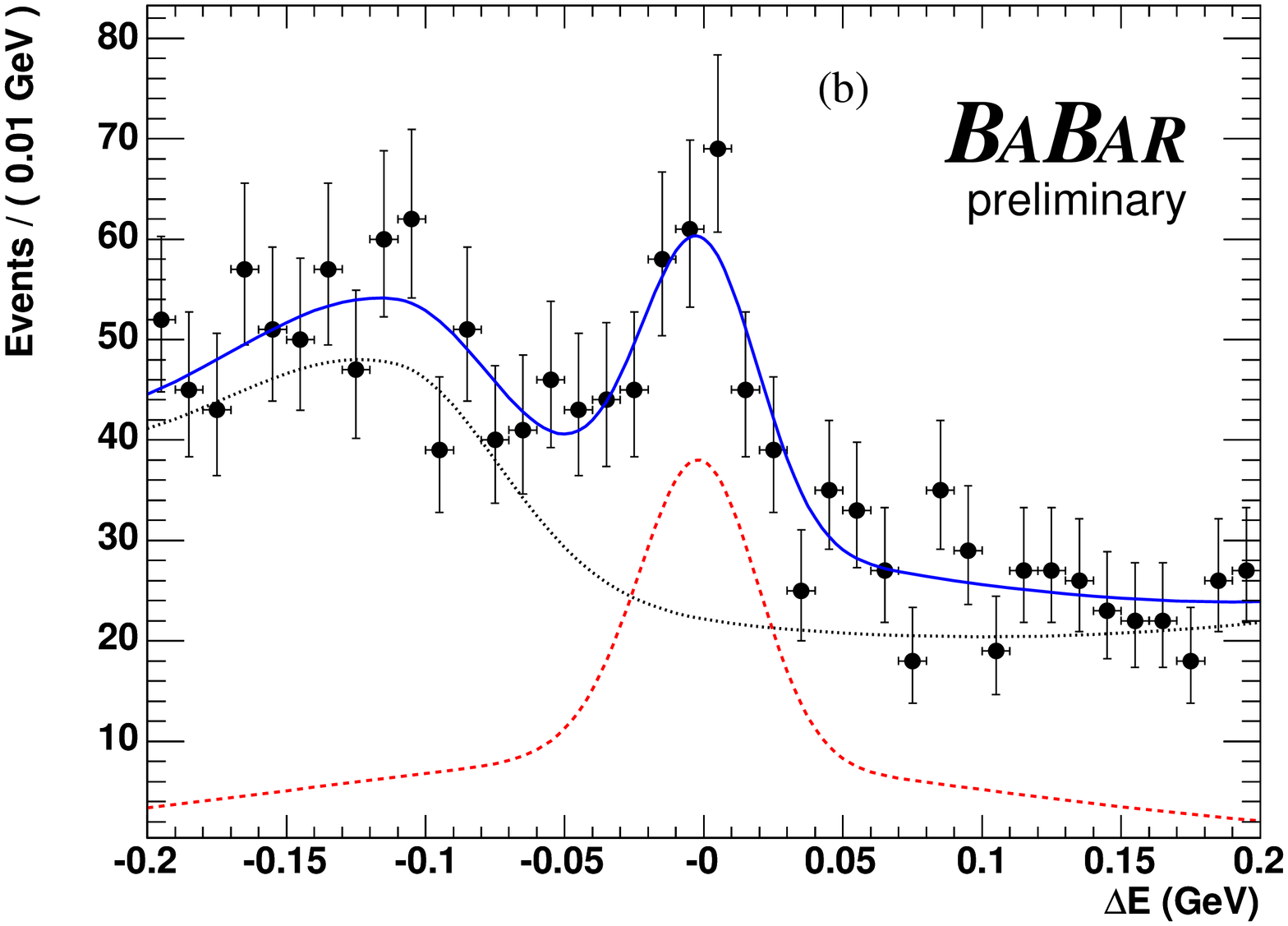}
\vspace{1.5cm}
 \hspace{3cm}
   \includegraphics[bb=85 155 535 605 ,angle=270,scale=0.30]{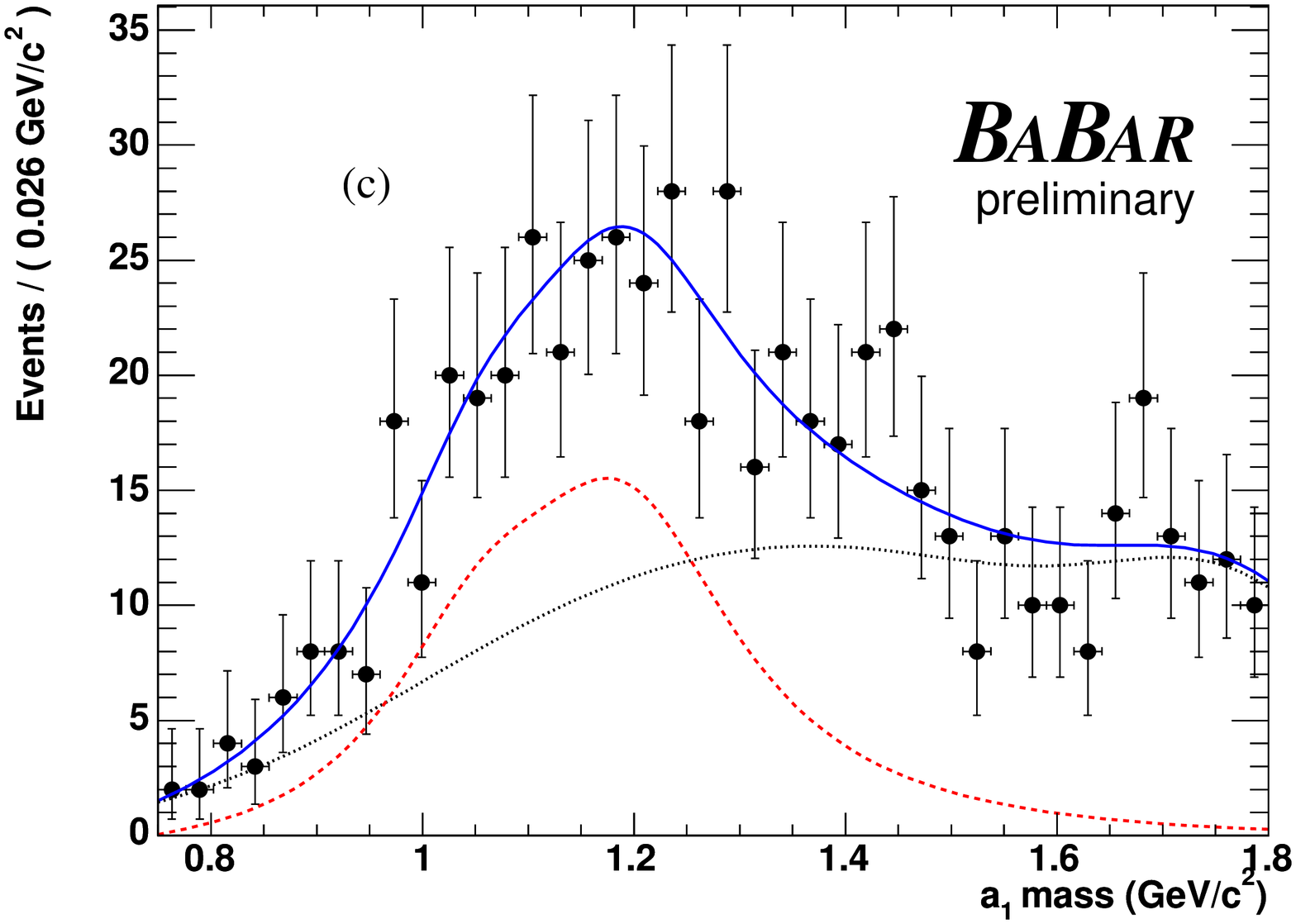}
\vspace{-1.5cm}
   \hspace{3cm}
 \vspace{0.5cm}
   \includegraphics[bb=85 155 535 605 ,angle=270,scale=0.30]{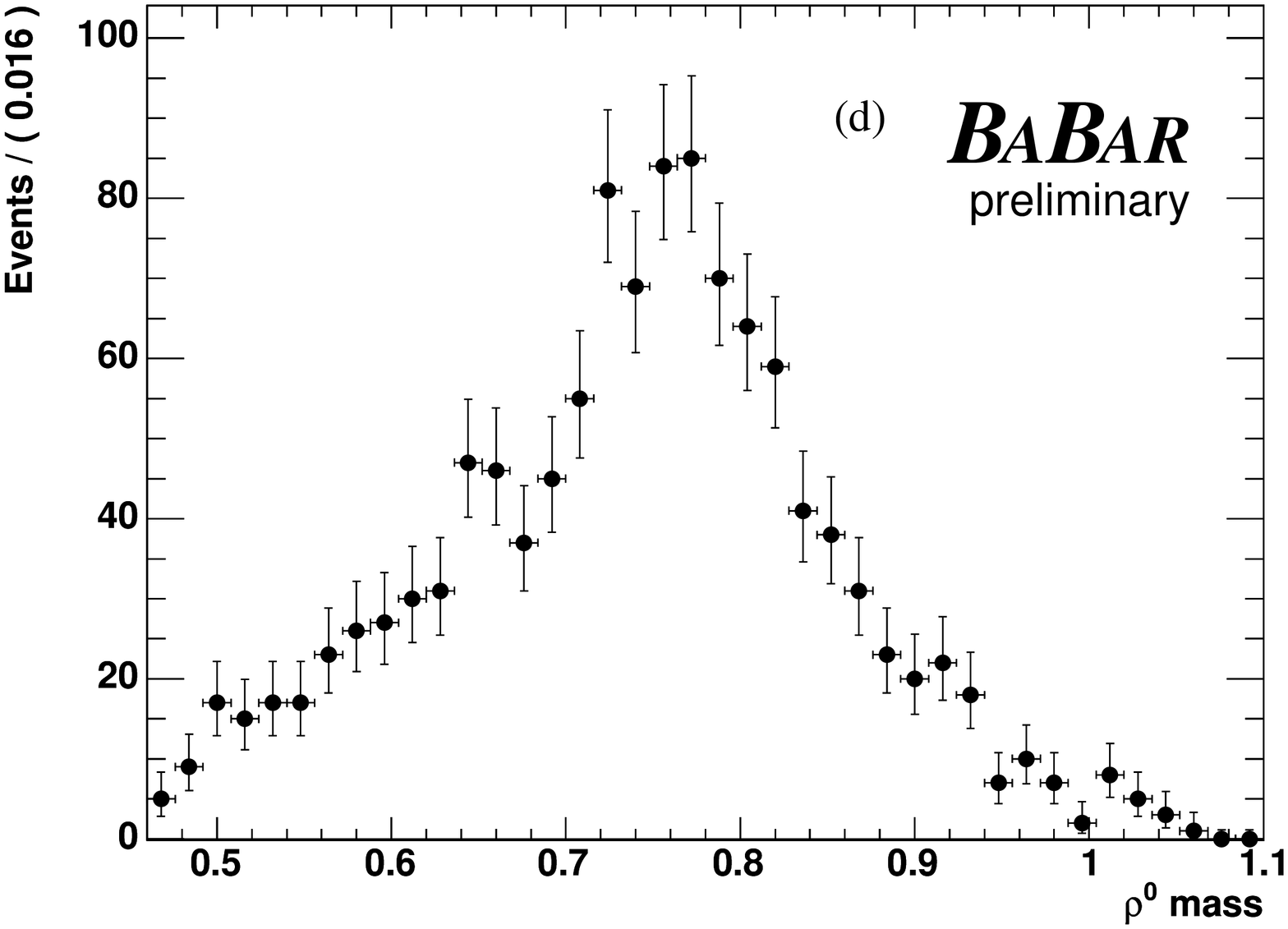}
\end{center}
  \end{minipage}
\vspace{1.cm}
  \caption{ Projections of \mes\ (a), \DE\ (b), $m_{a_1}$ (c), 
and  $\rho^0$ mass (d).
Points with errors represent data, dotted lines  the
continuum and \BB\ backgrounds, solid curves the full fit functions,
and dashed curves the signal.
These plots are made with a cut on the signal likelihood and thus do not
show all events in the data sample.}
\label{fig:projections}
\end{figure}

In Fig.\ \ref{fig:projections}\ we show the \mes, \DE, $m_{a_1}$,
and $m_{\rho}$ projections made by
selecting events with a signal likelihood (computed without the variable
shown in the figure) exceeding a threshold that optimizes the
expected sensitivity.

The fitted values of the \auno\ parameters are 
$m_{\aunob\ } = 1.19 \pm 0.02$ \gevcc\  and $\Gamma_{\aunob\ } = 312 \pm 55$ \mevcc\ .

\newpage
\section{SUMMARY}
\label{sec:Summary}
We have obtained a preliminary measurement of the branching fraction for $B^0$ meson decays
 to \appim\ with \auno\ $\rightarrow 3\pi$.
The measured  branching fraction is:

\begin{equation}
\Brapi = \Rapi,
\end{equation}

where the first uncertainty is statistical and the second uncertainty is systematic.
The fitted values of the \auno\ parameters are $m_{\aunob\ }=1.19 \pm 0.02$ \gevcc\ and
$\Gamma_{\aunob\ } = 312 \pm 55$ \mevcc. These values are closer to those found
in hadronic production of \auno\ meson.

\section{ACKNOWLEDGMENTS}
\label{sec:Acknowledgments}
We are grateful for the 
extraordinary contributions of our \pep2\ colleagues in
achieving the excellent luminosity and machine conditions
that have made this work possible.
The success of this project also relies critically on the 
expertise and dedication of the computing organizations that 
support \babar.
The collaborating institutions wish to thank 
SLAC for its support and the kind hospitality extended to them. 
This work is supported by the
US Department of Energy
and National Science Foundation, the
Natural Sciences and Engineering Research Council (Canada),
Institute of High Energy Physics (China), the
Commissariat \`a l'Energie Atomique and
Institut National de Physique Nucl\'eaire et de Physique des Particules
(France), the
Bundesministerium f\"ur Bildung und Forschung and
Deutsche Forschungsgemeinschaft
(Germany), the
Istituto Nazionale di Fisica Nucleare (Italy),
the Foundation for Fundamental Research on Matter (The Netherlands),
the Research Council of Norway, the
Ministry of Science and Technology of the Russian Federation, and the
Particle Physics and Astronomy Research Council (United Kingdom). 
Individuals have received support from 
CONACyT (Mexico),
the A. P. Sloan Foundation, 
the Research Corporation,
and the Alexander von Humboldt Foundation.


\begin{thebibliography}{99}
\bibitem{Coniugati}
 We use a particle name to denote either member of a charge conjugate pair.
\end{thebibliography}
\end{document}